\documentstyle[12pt]{article}
\newcommand{\al}{\alpha'}
\newcommand{\de}{\partial}
\newcommand{\be}{\begin{equation}}
\newcommand{\ba}{\begin{eqnarray}}
\newcommand{\ea}{\end{eqnarray}}
\newcommand{\ee}{\end{equation}}

\newcommand{\ca}{\mathcal}
\newcommand{\lr}{\leftrightarrow}
\newcommand{\f}{\frac}
\newcommand{\s}{\sqrt}

\newcommand{\ti}{\tilde}
\newcommand{\ap}{\alpha}

\newcommand{\ddd}{\cdot\cdot\cdot}
\newcommand{\no}{\nonumber \\}

\newcommand{\lb}{\rangle}

\begin{document}
\begin{titlepage}
\thispagestyle{empty}
\begin{flushright}
UT-02-36 \\
hep-th/0206010 \\
June, 2002
\end{flushright}

\bigskip

\begin{center}
\noindent{\large \textbf{Modular Invariance of 
Strings on PP-Waves with RR-flux
}}\\
\vspace{2cm}
Tadashi Takayanagi\footnote{takayana@hep-th.phys.s.u-tokyo.ac.jp}
\\
\vskip 2.5em

{\it Department of Physics, Faculty of Science, University of Tokyo\\
Hongo 7-3-1, Bunkyo-ku, Tokyo, 113-0033, Japan}

\vskip 2em

\end{center}

\begin{abstract}
We study the modular invariance of strings on pp-waves with RR-flux.
We explicitly show that the one-loop partition functions of the maximally 
supersymmetric
pp-waves and their orbifolds can be modular invariant in spite of the 
mass terms in the light-cone gauge. From this viewpoint, we also determine 
the spectrum of type 0B theory on pp-wave and discuss its gauge theory 
dual. Furthermore, we investigate the spectrum of a 
non-supersymmetric orbifold and
point out its supersymmetry enhancement in the Penrose limit.

\end{abstract}
\end{titlepage}

\newpage

\section{Introduction}
\setcounter{equation}{0}
\hspace{5mm}
Recently, much progress in understanding of
the superstring theory in RR background has been made 
originating from the exactly solvable plane-wave background 
with RR-flux \cite{Me}. Interestingly, this background possesses
an interpretation as the Penrose limit \cite{Pe} of the 
near horizon geometry 
of D3-branes \cite{BF,BeMaNa}.  The 
near horizon limit of $N$ D3-branes is given 
by $AdS_5\times S^5$ and its metric is written as 
\ba
ds^2=R^2\left(-dt^2\cosh^2 \rho+d\rho^2+\sinh ^2 \rho\ d\Omega^2_3
+d\psi^2\cos^2\theta+d\theta^2+\sin^2\theta d\Omega^{'2}_3\right),
\label{ads}
\ea
where the radius is given by $R=(4\pi g_sN\al^2)^{\f14}$.
After we take the Penrose limit $R\to\infty$ scaling as
$x^+=(t+\psi)/2\mu,\ x^-=R^2\mu (t-\psi), 
\ \rho\to\rho /R,\ \theta\to\theta /R$,
we obtain the maximally pp-wave metric \cite{BF,BeMaNa}
(with the constant RR-flux $F_{+1234}=F_{+5678}=\mu$)
\be
ds^2=-2dx^+dx^--\mu^2\sum_{i=1}^{8}(x^i)^2(dx^+)^2+\sum_{i=1}^{8}(dx^i)^2.
\label{ppwave}
\ee
Since we consider the RR background,
 we would like to quantize 
the world-sheet theory in the Green-Schwarz formulation of 
superstring \cite{GSW}. Remarkably,
the simple form of the metric (\ref{ppwave}) 
enables us to solve the world-sheet theory exactly in the 
light-cone gauge \cite{Me}.
Moreover, the exact string spectrum on pp-waves has been successfully
compared to the dual ${\ca{N}}=4$ gauge theory operators in \cite{BeMaNa}.
Further studies of string and M-theory in 
various pp-wave backgrounds and their
holographically dual relations
have been intensively carried out [6-30]\footnote{
In these developments the authors have discussed
the supersymmetry enhancement \cite{ItKlMu,GoOo,TaTa}, 
orbifolded pp-waves \cite{ItKlMu,AlJa,KiPaReTh,TaTe,FlKe,pporbifolde},
the less supersymmetric pp-waves \cite{RuTs,less-susy,Mi}, 
the algebras in the Penrose limit \cite{BF,susy-alg,HiSu}, 
D-branes in pp-waves
\cite{ppD-brane,TaTa,BeGaGr}, light-cone string field theory \cite{SpVo}, 
holographic
properties \cite{holography,BN}, 
string interactions in the gauge theory side \cite{BN,SYMInteraction}, 
the Penrose limit in NSNS background \cite{RuTs,TaTa,nsppwave,HiSu}, 
another
useful limit with a large value of spin \cite{GKP},
Matrix models \cite{Matrixmodel}
and some other gravitational properties \cite{gravity}.}.

However, the light-cone gauge theory has an important fault that
the conformal invariance is not manifest\footnote{The covariant string 
formalism of pp-wave background has been proposed in \cite{Be} and the 
conformal invariance has been checked.} because of the mass term 
$\sim(\mu p^+)^2\sum_{i}(X^i)^2$. At the same time, the modular invariance,
which is crucial for the quantum consistency of any string theory, is also
not clear. Thus we would like to investigate the modular 
invariance of string
theory on pp-waves in this paper. An earlier discussion of 
modular transformation of open-string amplitudes into boundary state 
amplitudes has been found in \cite{BeGaGr}. Below we will calculate
one-loop vacuum amplitudes of closed strings on pp-waves and 
examine their modular invariance. Since 
the pp-wave solution is the only solvable background with 
RR-flux at present,
the study of modular invariance will also be very important to know 
properties and consistencies of string spectra in general RR backgrounds.

The modular invariance also gives us a good principle to determine 
string spectra in new string backgrounds. As we will see below, 
we can determine the spectrum of various orbifolds of pp-waves from 
this viewpoint. The examples discussed in this paper 
include pp-wave in type 0B theory, the
non-supersymmetric oribifold ${\bf C}/{\bf Z}_M$ on pp-waves 
as well as the supersymmetric oribifold ${\bf C}^2/{\bf Z}_M$
theory on pp-waves \cite{ItKlMu,AlJa,KiPaReTh,TaTe,FlKe}. 
We also discuss the 
non-supersymmetric analog of the duality relation \cite{BeMaNa} between
the strings on pp-waves and the gauge theories
in the examples of type 0B and ${\bf C}/{\bf Z}_M$ pp-waves. 

This paper is organized as follows. In section 2, after a short review of
world-sheet theory of strings on pp-waves, we calculate the
one-loop vacuum amplitudes and examine their 
modular invariance. We apply this
method to type 0 theory and 
determine the spectrum of the type 0B pp-wave model. We also discuss the 
dual gauge theory in this model. In section 3, we construct 
modular invariant partition function in both supersymmetric and 
non-supersymmetric orbifolded pp-waves. We also discuss the gauge 
theory dual 
of a non-supersymmetric orbifold theory and its supersymmetry enhancement.
In section 4, we draw conclusions and present some discussions.

\section{Modular Invariance of Strings on PP-waves}
\setcounter{equation}{0}
\hspace{5mm}
Here we would like to investigate modular properties of string theory
on pp-waves. As we will see, we can prove its modular invariance 
performing a kind of
a deformation of ordinary arguments in string theory with no RR-flux.
See \cite{BeGaGr} for the analysis of modular transformation of 
open string amplitude into closed string one (Cardy's condition), which
is useful for the discussions below.

First we present a brief review of the Green-Schwarz string 
on pp-waves \cite{Me}. We mainly follow the convention of \cite{GSW,RuTs}.
The coordinate of world-sheet is represented by 
$\tau$ and $\sigma \ \ (0\leq \sigma \leq \pi)$.
After we take the light-cone gauge 
$X^+=2\al p^+\tau, 
\ \ \gamma^+\theta^{1,2}=0$, there are eight bosonic fields
$X^i\ \  (i=1\sim 8)$ and sixteen fermionic fields 
$S^{a},\ti{S^{a}}\ \ (a=1\sim 8)$. They are massive fields with the 
same mass
and the world-sheet 
action becomes 
\ba
S&=&\f{1}{\pi\al}\int d\tau d\sigma\left[\de_+X^i \de_- X^i-(\mu\al p^+)^2 
(X^i)^2\right]\no
&&+\f{i}{\pi}\int d\tau d\sigma\left[S^a\de_+S^a
+\ti{S}^a\de_- \ti{S}^a-(2\mu\al p^+)S^a\Pi^{ab}\ti{S}^b\right],
\ea
where we defined $\de_{\pm}=1/2(\de_{\tau}\pm\de_{\sigma})$ and 
$\Pi=\gamma^1\gamma^2\gamma^3\gamma^4$.

The mode expansions of the bosonic fields which satisfy 
the equation of motion
is given by
\ba
X^i\!=\!\cos(2\mu\al\! p^+\tau)x^i_{0}\!+\!
\f{\sin(2\mu\al\! p^+\tau)}{\mu p^+}
p^i_{0}\!+\! i\s{\f{\al}{2}}\!\!\sum_{n\neq 0}\!
\left(\!\f{\ap^i_{n}}{\omega_n}e^{-2i\omega_n\tau-2in\sigma}+
\f{\ti{\ap}^i_{n}}{\omega_n}
e^{-2i\omega_n\tau+2in\sigma}\!
\right)\!\!,
\ea
where we have defined 
\ba
\omega_{n}=\f{n}{|n|}\s{n^2+(\mu\al p^+)^2}.
\ea
The canonical quantization imposes the following commutation relations
\ba
[\ap^i_n,\ap^i_m]=[\ti{\ap}^i_n,\ti{\ap}^i_m]
=\omega_n\delta_{n+m,0}\delta^{ij},\ \ \ 
[\ap^i_0,\ap^{\dagger i}_0]=\omega_0,
\ea 
where we redefined the zero-modes 
as $\ap^i_{0}=\s{\al/2}(p^i_0-i\mu p^+ x^i_0),\
\ap^{\dagger i}_{0}=\s{\al/2}(p^i_0+i\mu p^+ x^i_0)$.

We can also perform the mode expansions of fermions 
$S^{a}$ and $\ti{S^{a}}$ 
similarly. We denote their oscillators as $S^a_{n}$ and $\ti{S}^{a}_n$
and (the linear combination of) the 
zero-modes as $S^a_{0}$ and $S^{\dagger a}_{0}$ such that the following 
anti-commutation relations are satisfied
\ba
\{S^a_{n},S^b_{m}\}=\{\ti{S}^{a}_n,\ti{S}^{b}_m\}=\delta_{n+m,0}\delta^{ab},
\ \ \ \{S^a_{0},S^{\dagger a}_{0}\}=\delta^{ab}.
\ea

Then we find the spectrum in the string theory on pp-wave as follows.
The light-cone Hamiltonian ${\ca{H}}=-p^-$ is given by
\ba
{\ca{H}}&=&\f{1}{\al p^+}\left(\ap^{\dagger i}_0
\ap^i_{0}+\omega_0 S^{\dagger a}_{0}S^a_{0}
+\sum_{n=1}^{\infty}(\ap^i_{-n} \ap^i_n+\ti{\ap}^i_{-n}\ti{\ap}^i_n
+\omega_n S^a_{-n}S^a_{n}+\omega_n \ti{S}^{a}_{-n}\ti{S}^{a}_{n})\right)\no
&\equiv&\sum_{n=-\infty}^{\infty}N_{n}
\s{\mu^2+\left(\f{n}{\al p^+}\right)^2}.\label{sp}
\ea 
The vacuum state $|0\lb$ has the zero light-cone energy $H|0\lb=0$ because
it is defined such that 
it is annihilated by the operators 
$\ap^i_{-n},\ti{\ap}^i_{-n},S^a_{-n},\ti{S}^{a}_{-n}\ \ (n>0)$ and 
$\ap^i_{0}$ and $S^a_{0}$. The level matching condition is also 
written as 
\ba
{\ca{P}}=\sum_{n=-\infty}^{\infty}nN_n=0.
\ea

Now the one-loop vacuum amplitude (partition function) 
can be defined as follows
\ba
Z=c\int\f{d\tau d\bar{\tau}}{\tau_2}\int dp^+dp^- \mbox{Tr}\left[
e^{-2\pi\al\tau_2 p^+ (p^-+{\ca{H}})+2\pi i\tau_1{\ca{P}}}\right],
\ea
where $\mbox{Tr}$ denotes the trace in the Hilbert space of the 
light-cone string theory\footnote{
For earlier discussions of closed string partition function
see \cite{RuTs}. Note that here we define the trace with the minus sign
for all spacetime 
fermions so that the amplitude represents the vacuum energy.}; 
$c$ is a constant factor, which we will 
 neglect below; $\tau=\tau_1+i\tau_2$ denotes the moduli of two 
dimensional torus as usual. The modular invariance of the 
partition function enables us to restrict the integration of the 
torus moduli $\tau,\bar{\tau}$ to the fundamental region.

\subsection{Modular Invariance of Type II String on PP-wave}
\hspace{5mm}
Before we compute the vacuum amplitude, 
it is useful to define the following partition function\footnote{
In ref.\cite{BeGaGr} the authors defined the real partition functions
$f_i^{(m)}(q),\ \ (i=1,2,3,4)$. Our complex partition function 
$Z^{(m)}_{a,b}(\tau,\bar{\tau})$ includes (the square of) them 
as the particular cases
$\tau_2=0$ and $(a,b)=(0,0),(0,1/2),(1/2,0),(1/2,1/2)$.}
$Z^{(m)}_{a,b}(\tau,\bar{\tau})$, which has a nice
modular property, 
\ba
Z^{(m)}_{a,b}(\tau,\bar{\tau})&=&e^{4\pi\tau_2\Delta^{(m)}_{b}}
\prod_{n=-\infty}^{\infty}(1-e^{-2\pi\tau_2\s{m^2+(n+b)^2}
+2\pi i\tau_1(n+b)+2\pi ia})\no
&&\ \ \ \ \ \ \ \times (1-e^{-2\pi\tau_2\s{m^2+(n-b)^2}
+2\pi i\tau_1(n-b)-2\pi ia}).
\ea
The factor $\Delta^{(m)}_{b}$ corresponds to 
the zero-energy (Casimir energy)
of a 2D complex scalar boson $\phi$ of mass $m$ with the 
twisted boundary condition 
$\phi(\tau,\sigma+\pi)=e^{2\pi ib}\phi(\tau,\sigma)$
and its explicit form is defined by
\ba
\Delta^{(m)}_{b}=-\f{1}{2\pi^2}\sum_{p=1}^{\infty}\int^{\infty}_{0} ds\ 
e^{-p^2s-\f{\pi^2m^2}{s}}\cos(2\pi bp).
\ea
In the massless limit this zero energy correctly reproduces the familiar
value
\ba
\lim_{m\to 0}\Delta^{(m)}_{b}=\f{1}{24}-\f{1}{8}(2b-1)^2.
\ea

The modular property of the above defined function is given by 
\ba
&&Z^{(m|\tau|)}_{a,b}(-\f{1}{\tau},-\f{1}{\bar{\tau}})=
Z^{(m)}_{-b,a}(\tau,\bar{\tau}), \label{MP}\\
&&Z^{(m)}_{a,b}(\tau+1,\bar{\tau}+1)=Z^{(m)}_{a+b,b}(\tau,\bar{\tau}).
\ea
The latter identity is easy to check and the more nontrivial relation 
(the former one) can be proved by using the Poisson resummation formula 
as shown in the appendix. The important point is that we must 
shift the value
of mass parameter as $m\to m|\tau|$. 
This is natural because the theory
is massive and thus is not conformal invariant.
Note also that since we consider the two dimensional 
massive theory, there is not any
definite distinction between left-moving and right-moving sector. Thus
we cannot define a sort of a chiral partition as in massless case.
It also enjoys other useful properties 
\ba
&&Z^{(m)}_{a+1,b}(\tau,\bar{\tau})
=Z^{(m)}_{a,b+1}(\tau,\bar{\tau})=Z^{(m)}_{a,b}(\tau,\bar{\tau}),\ \ \ 
Z^{(m)}_{-a,-b}(\tau,\bar{\tau})=Z^{(m)}_{a,b}(\tau,\bar{\tau}).
\ea
In the conformal limit, $m\to 0$ it is reduced to the familiar theta function
\ba
\lim_{m\to 0}Z^{(m)}_{a,b}(\tau,\bar{\tau})
\f{\s{Z^{(m)}_{0,0}(\tau,\bar{\tau})}}{2\pi\tau_2 m}=e^{-2\pi b^2\tau_2}
|\theta_{1}(a+b\tau|\tau)|^2. 
\ea

Now let us move on to the investigation of the one-loop string 
partition function on the maximally supersymmetric pp-wave. 
Its explicit form is given by\footnote{If we take the limit $\mu=0$, one may
think that the partition function should vanish due to Jacobi identity. 
This does not contradict our result (\ref{flatZ}) because our function 
$\sim 1/Z^{(m)4}_{0,0}$ includes the divergent volume factor.} 
\ba
Z_{\mbox{IIB}}=\int\f{d\tau d\bar{\tau}}{\tau_2}\int dp^+dp^-
e^{-2\pi\al\tau_2p^+ p^-}\left(
\f{Z^{(\al\mu p^+)}_{0,0}(\tau,\bar{\tau})}
{Z^{(\al\mu p^+)}_{0,0}(\tau,\bar{\tau})}\right)^4. \label{flatZ}
\ea 

Note that the denominator and numerator of the modular function 
are the same and thus the quotient is given by not zero but one\footnote{
After we submitted this paper, we were informed by A.A.Tseytlin 
of another definition
of one-loop vacuum amplitude based on path-integral 
calculations, which seems to be zero. 
The author is very grateful to A.A.Tseytlin for sending
his unpublished notes and our correspondence. In this paper we only 
discuss 
the partition function defined in the operator formalism.}. 
This phenomenon is due to the spacetime supersymmetry of the 
pp-wave background. 
Since the dynamical supercharges $Q^-$ commute with 
${\ca H}$ (see \cite{Me}), we can regard the total partition function
as the Witten index of $Q^-$. Thus only the (bosonic) 
ground state $|0\lb$ can contribute to it and we obtain the 
unit value\footnote{
The author would like to thank T. Eguchi and Y.Sugawara 
very much for related comments on the property of
partition function.}.

The modular invariance of (\ref{flatZ}) does hold formally without using 
this fact, 
that is, independently with respect to bosonic and fermionic 
contribution, though the integrations of $p^+$ and $p^-$ diverge.
 This is shown by noting the modular invariant combination
$d\tau d\bar{\tau}/\tau_2^2$ and performing the rescaling
\ba
p^{+'}=|\tau| p^+,\ \ \ \ \ p^{-'}=|\tau| p^-. \label{sca}
\ea
After we perform an 
analytic continuation of the momentum integration and 
integrate the torus moduli $\tau,\bar{\tau}$ in the fundamental 
region, we obtain
a finite value of the amplitude.

\subsection{Type 0 String on PP-wave}
\hspace{5mm}
Since in the previous example of the maximally supersymmetric pp-wave in 
type IIB theory
the modular function part is rather trivial due to the Bose-Fermi
degeneracy, next we would like to consider a more non-trivial case of
type 0B pp-wave\footnote{While preparing this paper for publication,
there appears the partially overlapped
paper
\cite{BiCoGiZa} which discusses string on type 0B pp-wave and 
duality to gauge theory
independently.}.

In type 0 string theory there are two types of
 D-branes each called electric and magnetic D-branes \cite{type0D} 
 (for a review of type 0 theory see \cite{Po}).
Since the near horizon limit of $N$ electric 
D3-branes and $N$ magnetic D3-branes in type 0B theory is given by 
$AdS_5\times S^5$ \cite{type0D},
the pp-wave background can be realized by considering the Penrose limit of 
this background. The gauge theory on these branes is non-supersymmetric and 
its gauge group is given by $SU(N)\times SU(N)$.
Generalizing the arguments in the ordinary type 0 theory 
with no RR-flux \cite{type0D},
we argue that the spectrum can be obtained by twisting that of 
type IIB pp-wave by the ${\bf{Z}}_2$ projection 
$(-1)^{F_S}$, where $F_S$ means the 
spacetime fermion number. Then we find the modular invariant 
partition function,
which is more complicated than in the type IIB case
\ba
Z_{\mbox{0B}}&=&\int\f{d\tau d\bar{\tau}}{\tau_2}\int dp^+dp^-
e^{-2\pi\al\tau_2p^+ p^-}\no
&& \times
\f{Z^{(\al\mu p^+)}_{0,0}(\tau,\bar{\tau})^4
+Z^{(\al\mu p^+)}_{1/2,0}(\tau,\bar{\tau})^4
+Z^{(\al\mu p^+)}_{0,1/2}(\tau,\bar{\tau})^4+
Z^{(\al\mu p^+)}_{1/2,1/2}(\tau,\bar{\tau})^4}
{2Z^{(\al\mu p^+)}_{0,0}(\tau,\bar{\tau})^4}. \label{flat0}
\ea 
The first two functions correspond to the untwisted sector of the 
${\bf{Z}}_2$ twist
and the last two correspond to the twisted sector with the shifted
modings as $\ap^i_{n+1/2},S^a_{n+1/2}$. We have determined their
non-trivial zero energy by requiring the modular invariance.
Note also that the spectrum is purely bosonic as in the flat background.

Next let us briefly discuss the duality of this pp-wave background 
to the gauge theory. This will give a non-supersymmetric generalization of 
the result in \cite{BeMaNa}. To see this it is useful to note that 
our system in type 0 theory can be obtained from a system of $2N$ D3-branes 
in 
type IIB theory ($SU(2N)$ gauge theory) 
by the projection $(-1)^{F_S}$. After the projection, open strings between
the same kind of D-branes are bosonic and open strings between
the different kind of D-branes (electric and magnetic) are fermionic.
More explicitly, the ${\bf Z}_2$ projection 
acts as\footnote{Here $\sigma_3$ is the Pauli matrix.
Below we omit the trivial part ${\bf{1}}_N$.} $(-1)^{F_S}
\cdot\sigma_3\otimes {\bf{1}}_N$ on the open string spectrum
with $2N\times 2N$ Chan-Paton matrices \cite{type0D}. 
Then we can see the duality to the
gauge theory by applying the arguments in orbifold theory 
\cite{AlJa,KiPaReTh,TaTe}.
The ground state of untwisted sector 
corresponds to the operator $\mbox{Tr}(Z^J)$, where
the trace is defined for $2N\times 2N$ matrices. The field 
$Z=\phi_{5}+i\phi_{6}$ 
denotes a (complex)
transverse scalar in the gauge theory and has the $U(1)$ charge $J=1$. 
The twisted sector
ground state is dual to the operator $\mbox{Tr}(\sigma_3 Z^J)$. 
The excitations of these vacuum states correspond to the insertions of 
the covariant derivatives $D_i\ \ (i=1\sim 4)$ or the other
transverse scalars $\phi^i\ \ (i=1\sim 4)$ taking the summation over their
possible positions 
with nontrivial phase factors in the same way as in \cite{BeMaNa}. Fermions
$\chi^a\ \ (i=1\sim 8)$, which have the $U(1)$ charge $J=1/2$,
can also be treated similarly.
For example, in the twisted sector we obtain the correspondence
\ba
S^a_{-n-1/2}\ti{S}^b_{-n-1/2}|0\lb\lr 
\sum_{l=0}^{J}e^{\f{2\pi il(n+1/2)}{J}}
\mbox{Tr}\left[\sigma_3Z^{l}\chi^aZ^{J-l}\chi^b\right],\label{ss}
\ea
where the twisted boundary condition
corresponds to the shift of the `momentum' $n$ by $1/2$ because of the 
non-trivial commutation relation 
$\chi^a\ \sigma_3=-\sigma_3\ \chi^a$ in the gauge theory as already
discussed in the context of orbifolded pp-waves \cite{KiPaReTh,TaTe}.
Interestingly, the spacetime fermions of type 0 string theory, which exist
only in open string sectors, just correspond to the purely bosonic
closed string spectrum via the holography. If we view this in an opposite
way, the absence of spacetime fermions in the closed string theory 
is consistent with the fact that the trace does vanish 
when we insert
the fermions into $\mbox{Tr}(Z^J)$ or $\mbox{Tr}(\sigma_3 Z^J)$ odd times.
It will also be interesting to perform gauge theory calculations and see
the matching of spectrum in detail including the non-trivial zero-energy
in the twisted sector (see (\ref{zero}) and later discussions).
We would like to leave further analyses for future problems.

Finally we would like to mention the 
closed string tachyon in type 0 theory. 
As is well-known,
there is a closed string tachyon 
in the flat background of type 0 string. In our case of pp-waves we can 
detect the tachyonic excitation by examining the divergent factor in
the partition function i.e. 
$\sim \int d\tau_2 e^{\tau_2\beta}\ \ (\beta>0)$. 
Thus we have only to know the difference of zero-energy for the latter two
functions in (\ref{flat0})
\ba
\Delta^{(\mu\al p^+)}_{1/2}-\Delta^{(\mu\al p^+)}_0=
\f{1}{\pi^2}\sum_{p=0}^{\infty}\int^{\infty}_{0} ds\ 
e^{-(2p+1)^2s-\f{(\pi\al\mu p^+)^2}{s}}>0. \label{zero}
\ea
This shows that we have always a tachyon for finite values of $\mu$.
If we set $\mu=0$, then the value becomes 
$\Delta^{(0)}_{1/2}-\Delta^{(0)}_0=1/8$ and the known tachyon mass 
$m^2=-\f{2}{\al}$ is reproduced. More interestingly, if we consider
the infinite value of $\mu$, then the tachyon seems to disappear.

\section{Modular Invariance of Orbifolded PP-Waves}
\setcounter{equation}{0}
\hspace{5mm}
The Green-Schwarz string on pp-wave background is originally constructed 
by employing its supersymmetries in the 
superspace formalism \cite{Me}. Thus it does not seem 
to be easy to obtain the less 
supersymmetric models in the Green-Schwarz formalism. However, the 
orbifold 
procedure allows us to realize such models with few
difficulties\footnote{We can also consider the Penrose limit of 
$AdS_3\times S^3\times \mbox{M}_{4}$ system 
\cite{BeMaNa,RuTs} as a less supersymmetric
example, where the four dimensional manifold ${\mbox{M}}_{4}$ 
represents $c=6$ 
conformal field theory without RR-flux.
The modular invariance of its partition function can be shown 
in the same way.}. We have already discussed the non-supersymmetric 
pp-wave in type 0 theory as a (non-geometric) orbifold and thus 
here we would 
like to investigate the orbifold defined by geometric projections.

The supersymmetric orbifolded pp-waves
has been considered and its duality to the quiver gauge theory \cite{DoMo}
has been checked including twisted sectors \cite{AlJa,KiPaReTh,TaTe}. 
In these arguments the authors impose the twisted boundary conditions
just as in the ordinary well-studied cases of orbifolds with no RR-flux.
Since the twisted sectors are originally added to physical sectors 
in order to keep
modular invariance (see e.g.\cite{Po}), 
it is very important to prove this property 
for orbifolded pp-waves.

\vspace{6mm}
{\sl Supersymmetric Orbifold ${\bf C^2}/{\bf Z}_M$}
\vspace{5mm}

Let us first discuss the orbifolded pp-waves which is equivalent to 
the Penrose limit of the near horizon geometry 
$AdS_5 \times S^5/{\bf Z}_M$
of $NM$ D3-branes 
at the orbifold ${\bf C^2}/{\bf Z}_M$. The orbifold action 
is given by
\ba
X=X^4+iX^5\mapsto e^{2\pi i/M}X,\ \ \ Y=X^6+iX^7\mapsto e^{-2\pi i/M}Y,
\ea
and the others $X^1,X^2,X^3,X^4$ are unchanged because they are in the 
direction of $AdS_5$ (only the coordinates of $\Omega_3^{'}$ in 
(\ref{ads}) 
are orbifolded). The fermions $S^a,\ti{S^a}$, which belong to the
spinor representation,
are also acted in the same way
because there are half of maximal spacetime supersymmetries preserved.
The partition function is written as follows
\ba
Z_{\mbox{orbs}}=\int\f{d\tau d\bar{\tau}}{\tau_2}\!\!\int dp^+dp^-
e^{-2\pi\al\tau_2p^+ p^-}\!\!\f{1}{M}\sum_{k,l=0}^{M-1} \left(
\f{Z^{(\al\mu p^+)}_{0,0}(\tau,\bar{\tau})}
{Z^{(\al\mu p^+)}_{0,0}(\tau,\bar{\tau})}\right)^2\!\!
\left(
\f{Z^{(\al\mu p^+)}_{k/M,l/M}(\tau,\bar{\tau})}
{Z^{(\al\mu p^+)}_{k/M,l/M}(\tau,\bar{\tau})}\right)^2\!\!\!, \label{orbs}
\ea
where the summations over $k$ and $l$ represent the orbifold projection and 
the twisted sectors, respectively.
We can show the modular invariance by using the property (\ref{MP}) only 
if we include the twisted sectors as in the above form (\ref{orbs}). 
Since 
there exist the partial supersymmetries preserved 
in this orbifolded pp-wave, the 
Bose-Fermi
degeneracy occurs and thus
the quotient of the modular functions is equal to a constant. We would also
like to comment that we can impose orbifold projections for all coordinates
$X^1\sim X^8$ in the same way, which will be the Penrose limit of the 
orbifold
not only with respect to 
$S^5$ but also to $AdS_5$.

\vspace{6mm}
{\sl Non-Supersymmetric Orbifold ${\bf C}/{\bf Z}_M$}
\vspace{5mm}

It will be also interesting to consider the non-supersymmetric orbifold of 
the maximally supersymmetric pp-wave\footnote{
In the paper \cite{TaTe} the non-supersymmetric type of the orbifold 
${\bf C^2}/{\bf Z}_M$ was considered. The discussion here can also be 
applied to this case without serious modifications.}. 
A simple example is the orbifold
${\bf C}/{\bf Z}_{M}$ ($M$ is an odd integer\footnote{Note that if 
we assume $M$ even, then it should be identified with 
an orbifold in type 0 theory \cite{AdPoSi,TaUe}.}) in pp-waves, 
where the orbifold projection
is defined by
\ba
X=X^4+iX^5\mapsto e^{2\pi iL/M}X,\label{za}
\ea 
where $L$ is an even integer (see e.g.\cite{c1,TaUe} 
for earlier discussions of its partition function without RR-flux).
Its modular invariant one-loop partition function is given by
\ba
Z_{\mbox{orbn}}\!=\!\int\f{d\tau d\bar{\tau}}{\tau_2}\!\int\! dp^+ 
\!dp^-\!
e^{-2\pi\al\tau_2p^+ p^-}\!\!\f{1}{M}\!\sum_{k,l=0}^{M-1} 
\f{\left(Z^{(\al\mu p^+)}_{\f{Lk}{2M},\f{Ll}{2M}}
(\tau,\bar{\tau})\right)^4}
{\left(Z^{(\al\mu p^+)}_{0,0}(\tau,\bar{\tau})\right)^3\!
Z^{(\al\mu p^+)}_{\f{Lk}{M},\f{Ll}{M}}(\tau,\bar{\tau})}. 
\label{orbn}
\ea 
The closed string tachyons in this model can also be examined as in the 
type 0 pp-wave and we find `localized tachyons' only in twisted sectors
as in the ordinary orbifold \cite{AdPoSi}. The duality to the 
non-supersymmetric
quiver gauge theory can also be examined as 
in the supersymmetric orbifold ${\bf C^2}/{\bf Z}_M$ formally, 
though
the matching of conformal dimension of operators is difficult to check 
due to the absence of BPS arguments. 
The $l$-th twisted sector corresponds to the
insertion of the diagonal matrix 
\ba
\gamma_{l}=\mbox{diag}
(1,e^{\f{2\pi iLl}{M}},e^{\f{4\pi iLl}{M}},\ddd,e^{\f{2\pi iL(M-1)l}{M}}),
\label{gam}
\ea
into the trace $\mbox{Tr}[Z^J]$.
It would also be intriguing to find the gauge 
theoretic origin of the non-trivial zero-energy 
$\Delta_{l}\equiv 
4\Delta^{(\al\mu p^+)}_{\f{Ll}{2M}}-3\Delta^{(\al\mu p^+)}_0
-\Delta^{(\al\mu p^+)}_{\f{Ll}{M}}>0$ in the $l$-th twisted sector. 

\vspace{6mm}
{\sl Supersymmetry Enhancement in Non-Supersymmetric Gauge Theory}
\vspace{5mm}

There is another Penrose limit of the previous non-supersymmetric 
orbifold. In the near horizon background (\ref{ads}) 
we can impose the ${\bf Z}_M$ projection
as follows
\ba
Z=R\cos\theta e^{i\psi}\mapsto e^{2\pi iL/M}\ Z.
\ea
After taking the limit, 
this leads to the periodicity in the light-cone direction 
(setting $\mu=1$)
\ba
(x^+ ,x^-)\sim (x^+ +\f{\pi L}{M},x^-+\f{2\pi LR^2}{M}),
\ea
where the radius is given by $R=(4\pi NMg_s\al^2)^\f{1}{4}$. 
When the $x^+$ direction is compactified, the background breaks maximal
supersymmetries  because the Killing spinor depends (only) 
on $x^+$ \cite{BF}.
If we take ,however, the limit $M,N\to\infty$ keeping $L$ and $\f{N}{M}$
finite, then we obtain the DLCQ compactification of maximally 
supersymmetric pp-waves (\ref{ppwave}) 
\ba
x^-\sim x^-+2\pi R^-\ \ \ (
R^-=2\al L\s{\pi g_sN/M}). \label{DLCQ}
\ea
This background preserves all of the 32 supersymmetries and thus the 
supersymmetry is enhanced 
in this limit even though the original background is 
non-supersymmetric\footnote{The enhancement of supersymmetry in the 
Penrose limit of non-supersymmetric models with NSNS-flux has also been 
discussed in \cite{TaTa}.}.
The similar type of the appearance of the DLCQ compactification has already 
been discussed in \cite{pporbifolde} in the case of the supersymmetric 
orbifold ${\bf C^2}/{\bf Z}_M$.

Let us consider the holographic relation between the string states of the 
DLCQ and the operators in the 
non-supersymmetric quiver gauge theory at the classical level
applying the arguments in \cite{pporbifolde}. To begin with, 
we obtain the following
correspondence 
\ba
p^-=\Delta-J,\ \ \  
2p^+=\f{\Delta+J}{R^2},
\ea
where $\Delta$ and $J$ are the conformal dimension and U(1) charge defined 
below, 
respectively. The light-cone compactification leads to the quantized 
momentum $p^+=\f{k}{R^-},\ \ (k\in {\bf Z})$. 
The light-cone Hamiltonian ${\ca{H}}(=-p^-)$ 
remains the same as (\ref{sp}) and
the level matching is given by
$\sum_{n=-\infty}^{\infty}nN_n=kw$, where $w$ is the winding number in the 
$x^-$ direction\footnote{The modular invariance of partition 
function in DLCQ
theory seems to be subtle because the scaling (\ref{sca}) is not allowed. 
This problem may suggest the requirement of the detailed analysis of 
Lorentzian torus. We leave this as a future problem.}. 

Next we would like to examine the quiver gauge theory.
There are six transverse scalars
and we denote them by $\phi^1,\phi^2,\phi^3,\phi^4$ and 
$Z=\phi^5+i\phi^6\ (\bar{Z}=\phi^5-i\phi^6)$. The ${\bf Z}_M$ projection
acts only on $Z$ and $\bar{Z}$ in the same as (\ref{za}). The fields 
which have non-zero R-charges are the scalars $Z\ \ (J=1)$, 
$\bar{Z}\ \ (J=-1)$ and fermions $\chi^1\sim \chi^8\ \ (J=1/2)$, 
$\ti{\chi}^1\sim \ti{\chi}^8\ \ (J=-1/2)$. We can expect that only 
the fields which have the value $\Delta-J=1$ will 
survive in the Penrose limit as in \cite{BeMaNa}. 
In order to be ${\bf Z}_M$
invariant the number of the insertions of $Z$ in any single trace operator
should be a multiple of $M$ (=$Mk$) and this just corresponds 
to the quantized
light-cone momentum $k$ in the DLCQ 
compactification (\ref{DLCQ}). Thus the vacuum state $|k,w=0\lb$ of DLCQ
string corresponds to the operator $\mbox{Tr}[Z^{Mk}]$, 
where the trace is defined 
for $NM\times NM$ matrices. Then the excited states can be obtained as
the insertions of covariant derivatives $D_i$, scalar fields $\phi^i$, or
fermions $\chi^a$. For example, we obtain the following 
translation for each insertion ($i=5,6,7,8$)
\ba
\ap^i_{-n}\lr \phi^i\ \gamma_{n/k},\ \ 
\ \ti{\ap}^i_{-n}\lr \phi^i\ \gamma_{-n/k},
\ea
where the matrix $\gamma$ is defined in (\ref{gam}). Note that after taking
the trace and summation over their positions $l$ (see (\ref{ss})), 
the insertions of extra matrices
$\gamma_{\pm n/k}$ are equivalent to both the familiar phase factor 
$e^{\pm\f{2\pi inl}{J}}$ and the overall matrix $\prod_n
(\gamma_{n/k})^{N_n}=\gamma_{m}$ in the first position in the trace.
The latter matrix comes from the orbifold action on Chan-Paton matrices
in the same 
way as before (see the arguments below (\ref{ss}) or \cite{KiPaReTh,TaTe}).

Even though the above arguments neglect quantum corrections, 
our `too good' results implies that the result will not be changed 
substantially if we go beyond the classical analysis\footnote{Here we 
ignore 
the presence of closed string tachyon in twisted sectors. Notice that the 
tachyonic instability seems to be reduced when we assume a large value of 
$\mu$ in the same way as in type 0 theory (\ref{zero}).}.
We would also like to mention
that the result suggests the possibility of `deconstructing' 
supersymmetric 
five dimensional gauge theory or six dimensional (2,0) theory from the
non-supersymmetric gauge theory (cf.\cite{dec,RoSk,pporbifolde}).

\section{Conclusions and Discussions}
\hspace{5mm}
In this paper we have investigated the modular invariance of strings
on pp-waves. At first sight, the models do not even seem to be
conformal invariant because of the mass terms for the world-sheet fields
in the light-cone gauge. However, the total theory including the 
light-cone directions $x^+,x^-$ should be completely conformal invariant 
as examined in the covariant formalism \cite{Be}. Our results show that the
 modular invariance can also be satisfied formally, 
 though the calculations include
 (perhaps inevitable) divergence due to Lorentz signature. 
 Thus we have obtained an important quantum consistency 
 of the string theory on pp-waves. It would be very interesting to see the 
 interpretation of the modular invariance and the scaling (\ref{sca})
 from the viewpoint of the dual gauge theory. 
 The one-loop contribution will be important when we consider non-planar
 diagrams in the gauge theories.

Since the modular invariance puts a rather strict constraint in string
theory, it is a useful guide when we construct a new theory. Based on this 
policy we have constructed the partition function of 
various orbifolded pp-waves. These include the supersymmetric 
orbifolds, non-supersymmetric type IIB orbifold and type 0B theory.
In the latter two examples, we also gave a brief survey of the 
correspondence between the strings on pp-waves
and dual (quiver) gauge theories. In particular, we pointed out that
supersymmetry can be enhanced in the specific Penrose limit
of the non-supersymmetric orbifold and the DLCQ compactification of
pp-waves can be obtained in a similar way to \cite{pporbifolde}.
We also find the possibility that in these examples
the tachyonic instability may be reduced for large values 
of RR-flux parameter. 

We believe the method developed in this paper will be useful in more 
general models.
For example, we will be able to discuss the world-sheet properties of 
fractional D-branes \cite{DoMo} in orbifolded pp-waves and also 
the modular properties of compactified pp-waves \cite{Mi}. Our results 
may also give a helpful hint when one tries to know an unknown 
NS-R formulation of strings on pp-waves with RR-flux.

\newpage
\noindent
{\bf Acknowledgments}

\vskip2mm
I am grateful to T.Eguchi and Y.Sugawara for useful
comments. I also thank A.A.Tseytlin for sending his unpublished 
notes and our correspondence.
The research of T.T.
was supported in part by JSPS Research Fellowships for Young
Scientists.

\appendix

\section{Detailed Calculations of Modular Transformations}
\setcounter{equation}{0}
\hspace{5mm}
Here we show the (deformed) modular transformation of 
$Z^{(m)}_{a,b}(\tau,\bar{\tau})$. 
In the following calculations we will only use the Poisson resummation 
formula
\ba
\sum_{n\in{\bf{Z}}}e^{-\pi an^2+2\pi inb}
=\f{1}{\s{a}}\sum_{\hat{n}\in{\bf Z}}e^{-\f{\pi}{a}(\hat{n}-b)^2},\label{pr}
\ea
and the identity
\ba
\f{1}{\s{\pi}}\int_0^{\infty}ds\ s^{-\f12}e^{-s-\f{t^2}{4s}}=e^{-t},
\label{inte}
\ea
generalizing the earlier computations done in \cite{BeGaGr}.

First we consider the logarithm of $Z^{(m)}_{a,b}(\tau,\bar{\tau})$
and perform the Poisson resummation
\ba
&&\log Z^{(m)}_{a,b}(\tau,\bar{\tau})-4\pi\tau_2\Delta^{(m)}_{b}\no
&&=\sum_{n\in{\bf Z}} \left[ \log(1-e^{-2\pi\tau_2\s{m^2+(n+b)^2}
+2\pi i\tau_1(n+b)+2\pi ia})
+\log(1-e^{-2\pi\tau_2\s{m^2+(n-b)^2}+2\pi i\tau_1(n-b)
-2\pi ia})\right]\no
&&=-\f{1}{\s{\pi}}\sum_{n\in\bf Z}\sum_{p=1}^{\infty}\int ds 
s^{-\f12}\left[
e^{-p^2 s-\pi^2\f{\tau_2^2}{s}(m^2+(n+b)^2)+2\pi i\tau_1 p(n+b)
+2\pi iap}+(\mbox{conjugate term})\right]\no
&&=-\f{1}{\pi\tau_2}\sum_{p=-\infty}^{\infty}
\sum_{\hat{n}=1}^{\infty}\int_0^{\infty}ds 
\Biggl[\exp\left(-\f{|\tau|^2}{\tau_2^2}
p^2 s+2\pi iap-\f{2\tau_1}{\tau_2^2}\hat{n}ps-\f{\pi^2\tau_2^2 m^2}{s}
-\f{s\hat{n}^2}{\tau_2^2}-2\pi ib\hat{n}\right)\no
&&+(\mbox{conjugate term})\Biggr]-4\pi\tau_2 \Delta^{(m)}_{b}+4\pi
\f{\tau_2}{|\tau|^2} 
\Delta^{(m|\tau|)}_{a}.\label{Mo1}
\ea

Further we again employ the Poisson resummation with respect to $p$ and
then the integral part of the final expression in (\ref{Mo1}) becomes
\ba
&&-\f{1}{\pi|\tau|}\sum_{\hat{p}=-\infty}^{\infty}
\sum_{\hat{n}=1}^{\infty}\int_0^\infty 
ds s^{-\f12}\exp\left(-\f{s\hat{n}^2}{|\tau|^2}
-\f{\pi^2\tau_2^2}{s}\left(\f{(\hat{p}\pm a)^2}{|\tau|^2}+m^2 \right)
\mp2\pi ibn\mp2\pi i(\hat{p}\pm a)\f{\tau_1\hat{n}}{|\tau|^2}\right)\no
&&=-\sum_{\hat{n}=1}^{\infty}\f{1}{\hat{n}}\sum_{\hat{p}=-\infty}^{\infty}
\exp\left(-2\pi\hat{n}\tau_2
\s{\f{(\hat{p}\pm a)^2}{|\tau|^4}+\f{m^2}{|\tau|^2}}\mp 2\pi 
i(\hat{p}\pm a)
\f{\tau_1\hat{n}}{|\tau|^2}\mp 2\pi ib\hat{n}\right).\label{Mo3}
\ea

Putting this result (\ref{Mo3}) into (\ref{Mo1}) we can easily find the 
modular transformation of $Z^{(m)}_{a,b}(\tau,\bar{\tau})$ (\ref{MP}).


\begin{thebibliography}{99}
\small
\baselineskip=11pt

\bibitem{Me}
R.~R.~Metsaev,
``Type IIB Green-Schwarz superstring in plane wave Ramond-Ramond
background,''
Nucl.\ Phys.\ B {\bf 625} (2002) 70, hep-th/0112044;
R.~R.~Metsaev and A.~A.~Tseytlin,
``Exactly solvable model of superstring in plane wave Ramond-Ramond
background,'' hep-th/0202109.

\bibitem{Pe}
R. Penrose, ``Any space-time has a plane wave as a limit,'' Differential
geometry and relativity, Reidel, Dordrecht, 1976 (271).

\bibitem{BF}
M.~Blau, J.~Figueroa-O'Farrill, C.~Hull and G.~Papadopoulos,
``A new maximally supersymmetric background of IIB superstring theory,''
JHEP {\bf 0201} (2002) 047, hep-th/0110242;
M.~Blau, J.~Figueroa-O'Farrill, C.~Hull and G.~Papadopoulos,
``Penrose limits and maximal supersymmetry,''
Class.\ Quant.\ Grav.\  {\bf 19} (2002) L87, hep-th/0201081;
M.~Blau, J.~Figueroa-O'Farrill and G.~Papadopoulos,
``Penrose limits, supergravity and brane dynamics,'' hep-th/0202111.

\bibitem{BeMaNa}
D.~Berenstein, J.~Maldacena and H.~Nastase,
``Strings in flat space and pp waves from N = 4 super Yang Mills,''
JHEP {\bf 0204} (2002) 013,
hep-th/0202021.

\bibitem{GSW}
M.~B.~Green, J.~H.~Schwarz and E.~Witten,
``Superstring Theory. Vol. 1,''
{\it  Cambridge, Uk: Univ. Pr.} (1987).

\bibitem{ItKlMu}
N.~Itzhaki, I.~R.~Klebanov and S.~Mukhi,
``PP wave limit and enhanced supersymmetry in gauge theories,''
JHEP {\bf 0203} (2002) 048,
hep-th/0202153.

\bibitem{GoOo}
J.~Gomis and H.~Ooguri,
``Penrose limit of N = 1 gauge theories,''
hep-th/0202157;
L.~A.~Zayas and J.~Sonnenschein,
``On Penrose limits and gauge theories,''
JHEP {\bf 0205} (2002) 010,
hep-th/0202186;
U.~Gursoy, C.~Nunez and M.~Schvellinger,
``RG flows from Spin(7), CY 4-fold and HK manifolds to AdS, Penrose 
limits and pp waves,''
hep-th/0203124;
C.~h.~Ahn,
``More on Penrose limit of AdS(4) x Q**(1,1,1),'' hep-th/0205008;
K.~Oh and R.~Tatar,
``Orbifolds, Penrose limits and supersymmetry enhancement,''
hep-th/0205067;
C.~h.~Ahn,
``Comments on Penrose limit of AdS(4) x M**(1,1,1),''
hep-th/0205109.

\bibitem{RuTs}
J.~G.~Russo and A.~A.~Tseytlin,
``On solvable models of type IIB superstring in NS-NS and R-R plane wave 
backgrounds,''
JHEP {\bf 0204} (2002) 021,
hep-th/0202179.

\bibitem{susy-alg}
M.~Hatsuda, K.~Kamimura and M.~Sakaguchi,
``From super-AdS(5) x S**5 algebra to super-pp-wave algebra,''
hep-th/0202190;
M.~Hatsuda, K.~Kamimura and M.~Sakaguchi,
``Super-PP-wave algebra from super-AdS x S algebras in 
eleven-dimensions,''
hep-th/0204002;
G.~Arutyunov and E.~Sokatchev,
``Conformal fields in the pp-wave limit,''
hep-th/0205270.

\bibitem{ppD-brane}
M.~Billo and I.~Pesando,
``Boundary states for GS superstrings in an Hpp wave background,''
hep-th/0203028;
C.~S.~Chu and P.~M.~Ho,
``Noncommutative D-brane and open string in pp-wave background 
with  B-field,'' hep-th/0203186; 
A.~Dabholkar and S.~Parvizi,
``Dp branes in pp-wave background,''
hep-th/0203231; 
D.~Berenstein, E.~Gava, J.~Maldacena, K.~S.~Narain 
and H.~Nastase,
``Open strings on plane waves and their Yang-Mills duals,''
hep-th/0203249; 
P.~Lee and J.~w.~Park,
``Open strings in PP-wave background from defect conformal 
field theory,''
hep-th/0203257;
A.~Kumar, R.~R.~Nayak and Sanjay,
``D-brane solutions in pp-wave background,'' hep-th/0204025;
D.~s.~Bak,
``Supersymmetric branes in PP wave background,''
hep-th/0204033;
K.~Skenderis and M.~Taylor,
``Branes in AdS and pp-wave spacetimes,'' hep-th/0204054;
V.~Balasubramanian, M.~Huang, T.~S.~Levi and A.~Naqvi,
``Open Strings from N=4 Super Yang-Mills,'' hep-th/0204196;
H.~Singh,
``M5-branes with 3/8 supersymmetry in pp-wave background,''
hep-th/0205020;
P.~Bain, P.~Meessen and M.~Zamaklar,
``Supergravity solutions for D-branes in Hpp-wave backgrounds,''
hep-th/0205106;
M.~Alishahiha and A.~Kumar,
``D-brane solutions from new isometries of pp-waves,''
hep-th/0205134;
C.~S.~Chu, P.~M.~Ho and F.~L.~Lin,
``Cubic string field theory in pp-wave background and background 
independent Moyal structure,''
hep-th/0205218;
S.~Seki,
``D5-brane in Anti-de Sitter Space and Penrose Limit,''
hep-th/0205266;
D.~Mateos and S.~Ng,
``Penrose Limits of the Baryonic D5-brane,''
hep-th/0205291;
S.~S.~Pal,
``Solution to Worldvolume Action of D3 brane in pp-wave Background,''
hep-th/0205303.

\bibitem{AlJa}
M.~Alishahiha and M.~M.~Sheikh-Jabbari,
``The PP-wave limits of orbifolded AdS(5) x S**5,''
hep-th/0203018.

\bibitem{KiPaReTh}
N.~w.~Kim, A.~Pankiewicz, S.~J.~Rey and S.~Theisen,
``Superstring on pp-wave orbifold from large-N quiver gauge theory,''
 hep-th/0203080.

\bibitem{TaTe}
T.~Takayanagi and S.~Terashima,
``Strings on orbifolded pp-waves,'' 
hep-th/0203093.

\bibitem{FlKe}
E.~Floratos and A.~Kehagias,
``Penrose limits of orbifolds and orientifolds,'' hep-th/0203134.

\bibitem{less-susy}
M.~Cvetic, H.~Lu and C.~N.~Pope,
``Penrose limits, pp-waves and deformed M2-branes,''
hep-th/0203082;
M.~Cvetic, H.~Lu and C.~N.~Pope,
``M-theory pp-waves, Penrose limits and supernumerary supersymmetries,''
hep-th/0203229;
J.~P.~Gauntlett and C.~M.~Hull,
``pp-waves in 11-dimensions with extra supersymmetry,''
hep-th/0203255;
H.~Lu and J.~F.~Vazquez-Poritz,
``Penrose limits of non-standard brane intersections,''
hep-th/0204001;
R.~Corrado, N.~Halmagyi, K.~D.~Kennaway and N.~P.~Warner,
``Penrose Limits of RG Fixed Points and PP-Waves with Background Fluxes,''
hep-th/0205314. 

\bibitem{Mi}
J.~Michelson,
``(Twisted) toroidal compactification of pp-waves,''
 hep-th/0203140.

\bibitem{gravity}
R.~Gueven,
``Randall-Sundrum zero mode as a Penrose limit,''
hep-th/0203153;
M.~Li,
``Correspondence principle in a pp-wave background,''
hep-th/0205043;
S.~D.~Mathur, A.~Saxena and Y.~K.~Srivastava,
``Scalar propagator in the pp-wave geometry obtained from AdS(5) x S**5,''
hep-th/0205136;
G.~Siopsis,
``Holography in the Penrose limit of AdS space,''
hep-th/0205302.

\bibitem{holography}
S.~R.~Das, C.~Gomez and S.~J.~Rey,
``Penrose limit, spontaneous symmetry breaking and holography in pp-wave  
background,'' hep-th/0203164;
E.~Kiritsis and B.~Pioline,
``Strings in homogeneous gravitational waves and null holography,''
hep-th/0204004;
R.~G.~Leigh, K.~Okuyama and M.~Rozali,
``PP-waves and holography,'' hep-th/0204026;
Y.~Imamura,
``Large angular momentum closed strings colliding with D-branes,''
hep-th/0204200.

\bibitem{Be}
N.~Berkovits,
``Conformal field theory for the superstring in a Ramond-Ramond plane 
wave background,''
JHEP {\bf 0204} (2002) 037, hep-th/0203248.

\bibitem{GKP}
S.~S.~Gubser, I.~R.~Klebanov and A.~M.~Polyakov,
``A semi-classical limit of the gauge/string correspondence,''
hep-th/0204051;
S.~Frolov and A.~A.~Tseytlin,
``Semiclassical quantization of rotating superstring in AdS(5) x S**5,''
hep-th/0204226;
J.~G.~Russo,
``Anomalous dimensions in gauge theories from rotating strings in 
$AdS_5\times S^5$,''
hep-th/0205244;
A.~Armoni, J.~L.~Barbon and A.~C.~Petkou,
``Orbiting strings in AdS black holes and N=4 SYM at finite temperature,''
hep-th/0205280.

\bibitem{SpVo}
M.~Spradlin and A.~Volovich,
``Superstring Interactions in a pp-wave Background,''
hep-th/0204146.

\bibitem{pporbifolde}
S.~Mukhi, M.~Rangamani and E.~Verlinde,
``Strings from Quivers, Membranes from Moose,''
hep-th/0204147;
M.~Alishahiha and M.~M.~Sheikh-Jabbari,
``Strings in PP-Waves and Worldsheet Deconstruction,''
hep-th/0204174.

\bibitem{TaTa}
H.~Takayanagi and T.~Takayanagi,
``Open strings in exactly solvable model of curved space-time and  
PP-wave limit,''
JHEP {\bf 0205} (2002) 012, hep-th/0204234.

\bibitem{nsppwave}
I.~Bakas and K.~Sfetsos,
``PP-waves and logarithmic conformal field theories,''
hep-th/0205006;
A.~Parnachev and D.~A.~Sahakyan,
``Penrose limit and string quantization in AdS(3) x S**3,''
hep-th/0205015;
V.~E.~Hubeny, M.~Rangamani and E.~Verlinde,
``Penrose Limits and Non-local theories,''
hep-th/0205258.

\bibitem{BN}
D.~Berenstein and H.~Nastase,
``On lightcone string field theory from super Yang-Mills and holography,''
hep-th/0205048.

\bibitem{SYMInteraction}
C.~Kristjansen, J.~Plefka, G.~W.~Semenoff and M.~Staudacher,
``A new double-scaling limit of N = 4 super Yang-Mills theory and 
PP-wave  
strings,''
hep-th/0205033;
D.~J.~Gross, A.~Mikhailov and R.~Roiban,
``Operators with large R charge in N = 4 Yang-Mills theory,''
hep-th/0205066;
N.~R.~Constable, D.~Z.~Freedman, M.~Headrick, S.~Minwalla, L.~Motl, 
A.~Postnikov and W.~Skiba,
``PP-wave string interactions from perturbative Yang-Mills theory,''
hep-th/0205089;
R.~Gopakumar,
``String interactions in PP-waves,''
hep-th/0205174;
Y.~j.~Kiem, Y.~b.~Kim, S.~m.~Lee and J.~m.~Park,
``PP-wave/Yang-Mills Correspondence: An Explicit Check,''
hep-th/0205279;
M.~-x.~Huang,
``Three point functions of N=4 Super Yang Mills from light cone string 
field theory in pp-wave,''
hep-th/0205311.

\bibitem{BeGaGr}
O.~Bergman, M.~R.~Gaberdiel and M.~B.~Green,
``D-brane interactions in type IIB plane-wave background,''
hep-th/0205183.

\bibitem{Matrixmodel}
K.~Dasgupta, M.~M.~Sheikh-Jabbari and M.~Van Raamsdonk,
``Matrix perturbation theory for M-theory on a PP-wave,''
hep-th/0205185;
G.~Bonelli,
``Matrix strings in pp-wave backgrounds from deformed super 
Yang-Mills  theory,''
hep-th/0205213.

\bibitem{HiSu}
Y.~Hikida and Y.~Sugawara,
``Superstrings on PP-wave backgrounds and symmetric orbifolds,''
hep-th/0205200.

\bibitem{BiCoGiZa}
F.~Bigazzi, A.~L.~Cotrone, L.~Girardello and A.~Zaffaroni,
``PP-wave and Non-supersymmetric Gauge Theory,''
hep-th/0205296 

\bibitem{type0D}
I.~R.~Klebanov and A.~A.~Tseytlin,
``D-branes and dual gauge theories in type 0 strings,''
Nucl.\ Phys.\ B {\bf 546} (1999) 155, hep-th/9811035;
I.~R.~Klebanov and A.~A.~Tseytlin,
``A non-supersymmetric large N CFT from type 0 string theory,''
JHEP {\bf 9903} (1999) 015, hep-th/9901101.

\bibitem{Po}
J.~Polchinski,
``String Theory. Vol. 1,2,''
{\it  Cambridge, UK: Univ. Pr.} (1998).

\bibitem{DoMo}
M.~R.~Douglas and G.~W.~Moore,
``D-branes, Quivers, and ALE Instantons,''
hep-th/9603167;
D.~Diaconescu, M.~R.~Douglas and J.~Gomis,
``Fractional branes and wrapped branes,''
JHEP{\bf 9802} (1998) 013,
hep-th/9712230.

\bibitem{AdPoSi}
A.~Adams, J.~Polchinski and E.~Silverstein,
``Don't panic! Closed string tachyons in ALE space-times,''
JHEP {\bf 0110} (2001) 029, hep-th/0108075.

\bibitem{TaUe}
T.~Takayanagi and T.~Uesugi,
``Orbifolds as Melvin geometry,''
JHEP {\bf 0112} (2001) 004, hep-th/0110099.

\bibitem{c1}
A.~Dabholkar,
``Strings on a cone and black hole entropy,''
Nucl.\ Phys.\ B {\bf 439} (1995) 650,hep-th/9408098;
D.~A.~Lowe and A.~Strominger,
``Strings near a Rindler or black hole horizon,''
Phys.\ Rev.\ D {\bf 51} (1995) 1793, hep-th/9410215.

\bibitem{dec}
N.~Arkani-Hamed, A.~G.~Cohen and H.~Georgi,
``(De)constructing dimensions,''
Phys.\ Rev.\ Lett.\  {\bf 86} (2001) 4757, hep-th/0104005;
N.~Arkani-Hamed, A.~G.~Cohen, D.~B.~Kaplan, A.~Karch and L.~Motl,
``Deconstructing (2,0) and little string theories,''
hep-th/0110146.

\bibitem{RoSk}
I.~Rothstein and W.~Skiba,
``Mother moose: Generating extra dimensions from simple groups at 
large  N,''
Phys.\ Rev.\ D {\bf 65} (2002) 065002, hep-th/0109175.

\end{thebibliography}
\end{document}